\documentclass[11pt]{article}

\usepackage[utf8]{inputenc}
\usepackage[english]{babel}
\usepackage{array}
\usepackage{amsmath,amssymb}
\usepackage{graphics}
\usepackage{graphicx}
\usepackage{color}

\usepackage{amsthm}

\usepackage{bm} 

\usepackage{setspace}

\newtheorem{theo} {Theorem}

\newtheorem{propo} {Proposition}

\newtheorem{defi}{Definition}
\newtheorem{rem}{Remark}
\newtheorem{nota}{Notations}
\newtheorem*{nota*}{Notations}

\newtheorem*{ex*}{Example}

\newcommand{\cqfd}{\hfill $\Box$ \vskip 0.5cm}

\def \zero {\underline{0}}
\def \un {\underline{1}}
\def \AA {\mathbb{A}}
\def \FF {\mathbb{F}}
\def \II {\mathbb{I}}
\def \LL {\mathbb{L}}
\def \OO {\mathbb{O}}
\def \RR {\mathbb{R}}
\def \SS {\mathbb{S}}

\def \id {\mathsf{id}}
\def \pardef {\stackrel{\mathsf{def}}{=}}
\def \split {\bot}
\def \cop {\parallel}

\def \emstate {\hat{\mathcal{E}}}
\def \Setemstates {\hat{E}}
\def \attractor {\hat{\mathcal{E}}^{att}}
\def \empaths {\mathcal{E}}
\def \fhat {\hat{f}}
\def \Setpaths {{\cal P}}
\usepackage[mathcal]{euscript}

\DeclareMathOperator*{\argmax}{arg\,max}

\title{Self-organization and the Maximum Empower Principle in the Framework of max-plus Algebra}

\author{Chams LAHLOU\footnote{chams.lahlou@imt-atlantique.fr} \ and Laurent TRUFFET\footnote{laurent.truffet@imt-atlantique.fr}\\
Department of  Automation, Production and Computer Sciences,\\
IMT-A, Nantes, France
}

\begin{document}

\maketitle

\begin{abstract}
Self-organization is a process where order of a whole system arises
out of local interactions between small components of a system.
\\
Emergy, spelled with an 'm', defined as the amount of (solar) energy
used to make a product or service, is becoming an important ecological
indicator. The Maximum Empower Principle (MEP) was proposed as the fourth
law of thermodynamics by the ecologist Odum in the 90's to explain observed self-organization of energy driven
systems. But this principle
suffers a lack of mathematical formulation due to an insufficiency of
details about the underlying computation of empower (i.e. emergy per
time).
\\
For empower computation in steady-state an axiomatic basis has been
developed recently by Le Corre and the second author of this paper. In
this axiomatic basis emergy is defined as a recursive max-plus linear
function.
\\
Using this axiomatic basis and a correspondance between ecological theory and dynamic systems theory, we prove the MEP. In particular, we show that the empower computation in steady-state is equivalent to a combinatorial optimization problem.
\end{abstract}      


\section{Introduction}
It has been observed since a long time (see e.g. \cite{kn:Podolinsky1880},
\cite{kn:Boltzmann1886}) that energy, as the ability to
do work, plays an important role in our civilization. Nowadays, more and more
people realize that complex systems such as ecological networks, social
organizations, economic systems are energy driven systems. 

Self-organization, or spontaneous order principle, states that any
living or non-living disordered system evolves towards an “equilibrium
state” or coherent state, also called attractor. Self-organization is
observed e.g. in physical systems (\cite{kn:Bar-Yam97},
\cite{kn:Haken78}, \cite{kn:Nico77}), in biological systems
\cite{kn:Camaetal03}, in social systems \cite{kn:Anzolaetal2017}, in
mathematical systems/models, in economics, in information theory and
informatics (see e.g. \cite{kn:Kaufman93} and references therein).

To explain self-organization of energy driven systems, the maximum power
principle has been proposed in e.g. \cite{kn:Lotka22}
and \cite{kn:Odum55}. This principle states that: \\
“system designs develop and prevail that maximize power intake, energy
transformation, and those uses that reinforce production and
efficiency”.

The major drawback of such approach is that complex energy systems can
use energies of different kinds, e.g. renewable energies (solar, wind, ...)
fossiles energies (fuel, gaz, coal), nuclear energy. Moreover, different
energies do not have the same time scale. In \cite[Chap. 2]{kn:Odum96}
the concept of energy hierarchy is introduced. It means that if the Sun
is the reference point and is considered to be instantaneously available, then
e.g. the fuel requires thousands of years to be used by human beings. And
these two energies do not have the same calorific power.

In order to address this problem the ecologist Odum proposed the
concept of emergy (spelled with an 'm' which is a neologism for energy
memory). This term was coined in the
mid-80's in e.g. \cite{kn:Scienceman87}. In \cite[p. 7]{kn:Odum96} it is defined as follows: “Emergy is defined as the
available energy of one kind previously used up directly or indirectly
to make a service or a product". It is a 
cumulative function of available energy and its unit is the
emjoule. Recalling as abovementioned that different kinds of energies
do not have the same ability to do work, Odum proposed to take the Sun
as the reference point and defined the solar emergy as the available
solar energy used directly or indirectly to make a service or product.
Its unit is the solar emjoule, abbreviated sej \cite[p. 8]{kn:Odum96}.
Thus, solar emergy can be considered as a metric for environmental
assessment which allows to compare different energy systems doing
the same functions on the same basis: the Sun.

The major contributions of Odum are: \\

{\bf Transformity}. To take into account the different time scale of 
energies, Odum introduced 
the dimensionless number he called transformity. The transformity is defined
as the emergy required to make one Joule of a service or product
\cite[p.10, p. 288]{kn:Odum96}, so that we have:
\[
\mbox{emergy} \pardef \mbox{transformity} \times \mbox{available energy}.
\]

{\bf Process path function}. Emergy of a product or service is a 
function of solar energy and its value depends on the scenario 
followed by the solar energy to generate the product or service 
under examination.

{\bf Maximum Empower Principle (MEP)}. 
Defining the empower as the emergy per time Odum proposed the maximum
empower principle (MEP) to explain self-organization of energy networks
as a Universal principle (fourth law of thermodynamics). \\
{\bf MEP:} ''In the
competition among self-organizing processes, network designs that maximize
empower will prevail'' \cite[p. 16]{kn:Odum96}. A network design that
maximizes empower is named a sustainable design \cite[p. 279]{kn:Odum96}. 

The concept of emergy as an holistic paradigm which allows to compare
two energy systems on the same basis (i.e. solar emergy) has generated a
lot of literature on the subject and has been successfully applied on
many domains (see e.g.  \cite{kn:Chenetal2017} and references
therein).

But the concept of emergy has also generated debates and criticisms (see
e.g.  \cite{kn:Hau2004} and references therein). As mentioned in 
\cite[Sec. 3.2]{kn:Hau2004}: ''it is important to note that many 
criticisms are also valid for other methods [...] including 
Life Cycle Assessment, Cumulative Exergy analysis, ...''. 

However, the major drawback of the empower computation was a 
{\bf lack of mathematical formalism}. 
Assuming the following hypothesis (see Section~\ref{secRecall} for details):

\begin{itemize}
\item(A0) Steady-state analysis.
\item(A1) No creation of emergy.
\item(A2) Emergy in feedbacks cannot be added more than once,
\end{itemize}
an answer to the challenging problem of computing empower or 
emergy through complex networks was proposed in \cite{kn:Lecorre2012b}. 
In this framework the $(\max,+)$-algebra or tropical algebra 
(see e.g. \cite{kn:Bac-cooq}) plays a central role. The algorithm 
provided by the axiomatic basis developed in \cite{kn:Lecorre2012b} 
has been successfully applied in \cite{kn:Lecorre2012}, and in 
\cite{kn:Lecorre2015} (complex farm analysis).

\subsection{Contribution of the paper}
The maximum empower principle as the fourth principle of thermodynamics 
has received criticisms (see e.g. \cite{kn:Mansson93}) and 
rebuttals (see e.g. \cite{kn:Odum95}, \cite{kn:Odum02}, 
\cite{kn:LiLuetal13} ) since it was stated. 

Under:
\begin{itemize}
\item assumptions (A0)-(A2)
\item axiomatic basis developed in \cite{kn:Lecorre2012b}
\end{itemize}
the maximum empower principle is proved (see Theorem~\ref{theo:maxempower}).

It is important to notice that our result does not depend on 
the exact definition of available energy. It is just implicitly 
assumed that it is a nonnegative quantity linked to energy 
concepts.

\subsection{Related works}
To the best knowledge of the authors only one pioneering 
work concerning the mathematical formulation of the MEP 
was developed in \cite{kn:Giannantoni02}. This work is 
based on:

\begin{itemize}
\item assumptions (A1)-(A2),

\item linear algebra,

\item fractional calculus,

\item available energy defined as exergy.

\end{itemize}

The framework of \cite{kn:Giannantoni02} is more general than the one 
of this paper however we can make the following two remarks:

Linear algebra is not the appropriate framework for emergy computation. 
Indeed, it has been noticed in e.g. \cite{kn:Patterson2014} that 
this approach can lead to absurd results such as negative transformities. 

The emergy is defined as the space and time integral of the exergy but, 
in fact, the Gibb's free energy is used (see 
\cite[p. 3700, footnote 4]{kn:Sciubba10}).

\subsection{Organization of the paper}

First, we introduce in Section \ref{secRecall} two important notions, which are \textit{emergy graph} and \textit{emergy path}, then we recall the axiomatic basis (developed in \cite{kn:Lecorre2012b}) on which the MEP is proved.

Then, in Section \ref{secMEP}, we present the correspondance between ecological theory and dynamic systems theory (see Table \ref{tab:correspondance}). Using this correspondance we establish the MEP (see Theorem \ref{theo:maxempower}).

Section \ref{secExample} is devoted to a numerical example which illustrates all the concepts developed in the paper.

Finally, in Section \ref{secConclusion}, we reformulate the MEP using our settings and suggest a new line of approach for empower computation.

\section{Emergy calculus reminder}
\label{secRecall}

In this section we recall basic materials to compute empower or emergy
through networks. A network is modelled by a particular valued
directed graph named in the sequel emergy graph which is a multiple
inputs multiple outputs system. Let us recall and detail our main three
assumptions.

\begin{itemize}
\item[(A0)] Steady-state analysis. It means that the 
characteristics of the emergy graph (topology, valuation) 
does not depend on the time. 

\item[(A1)] No creation of emergy. The emergy received by an output 
  cannot be greater than the emergy of the input from which it is
  derived.

\item[(A2)] Emergy in feedbacks cannot be added more than once. The 
only paths in the emergy graph which are of importance are 
the emergy paths (see Definition~\ref{defPath}). Emergy path 
is either a particular case of simple path or terminal 
non-feedback cycle path, which are well-known in ecology 
(see e.g. \cite{kn:Whipple99}).
\end{itemize}

\subsection{Emergy graph}
\label{subsec:emgraph}

\begin{figure}
\centering
\includegraphics[scale=0.5]{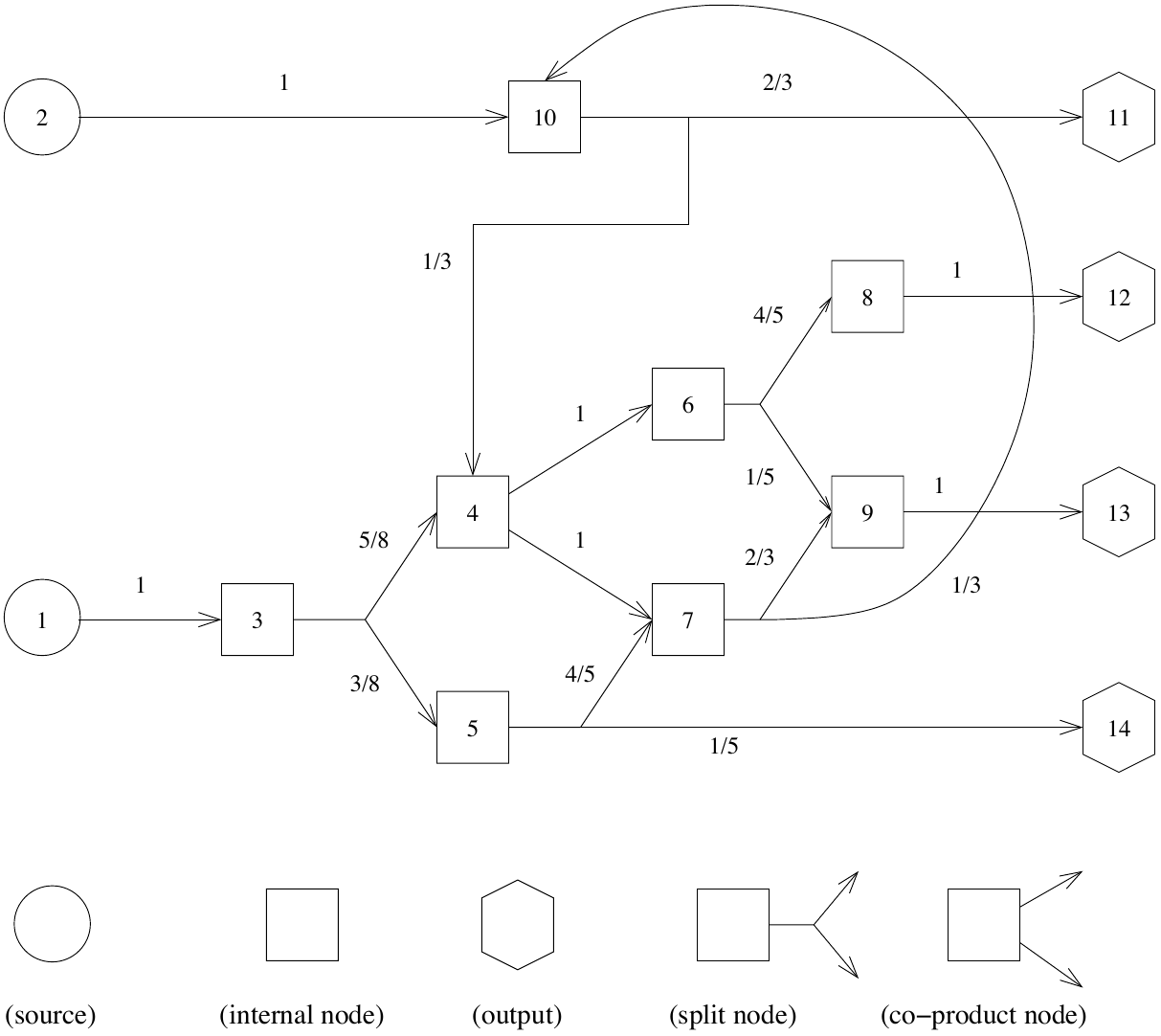}
\caption{An emergy graph with five splits and one co-product at node 4 \cite{kn:Lietal2010}.}
\label{fig:example}.
\end{figure}

The way by which emergy circulates in a multicomponent system is
modelled by a directed graph, which is called emergy graph
\cite{kn:Lecorre2012b}. An example, taken from \cite{kn:Lietal2010}, is given in Figure \ref{fig:example}. Formally, it is the following 10-tuple:
	 	
$$
G \pardef (\LL, \SS, \II, \OO, F, \AA, \id, \split, \cop, \emptyset).
$$

An emergy graph has different kinds of nodes stored in $\LL$.

\begin{itemize}

\item The nodes which characterize the boundaries of the system:

\begin{itemize}
  
\item The source nodes stored in $\SS$. These nodes are the inputs of
  the system. They are
  associated with different kind of energies (renewable, fossile, nuclear).
  Their emergy is defined by the emergy 
  function $\theta$ (see definition in subsection~\ref{subEmFct}). In Figure \ref{fig:example} we have $\SS =\{1,2\}$.

\item The product nodes stored in $\OO$. These nodes are the outputs of
  the system. In Figure \ref{fig:example} we have $\OO =\{11,12,13,14\}$

\end{itemize}

\item The nodes within the system stored in $\II$. For the example, $\II =\{3, 4, 5, 6, 7, 8, 9, 10\}$.
We distinguish:

  \begin{itemize}

  \item The split nodes. At a split node the available energy divides into available energy
    of one kind. Thus, the transformities downstream this node are equal.
    The emergy is assigned to each arc downstream the split proportionaly
    to the available energy flowing on the arc \cite[p. 91]{kn:Odum96}. This
    proportionality is modelled by the $[0,1]$-valued weight function
    $\omega$ (see definition in subsection~\ref{subEmFct}). 
    For the example, the split nodes are 3, 5, 6, 7, 8, 9 and 10.

  \item The co-product nodes. At a co-product node the available energy divides
    into available energies of different kinds on each arc after the co-product
    node e.g. as in combined heat and power plants (described in
    e.g. \cite{kn:Horlock96}). This means that transformities on each
    arc are different. And emergy has the same value on each arc
    downstream the co-product node than the emergy upstream the co-product
    node, i.e. the weight function $\omega$
    is $1$ on each arc downstream the co-product node.
    In the example there is only one co-product node, which is node 4.

  \end{itemize}

\end{itemize}
The sets $\SS$, $\II$ and $\OO$ form a partition of the set $\LL$.

The arcs of the graph are stored in $\AA$. An arc between the two
processes $l_1$ and $l_2$ is denoted 
$[l_1 ; l_2]$. It represents the fact that emergy can flow from 
$l_1$ to $l_2$. 

The set $\AA$, $\SS$ and $\OO$ satisfy
$$
\AA \cap (\SS \times \SS) = \emptyset,
$$
which means that two different sources cannot be linked, and
$$
 \AA \cap (\OO \times \OO)= \emptyset,
$$
which means that two different outputs of the system cannot be linked.

Every pair of arcs of an emergy graph must satisfy one of the four
binary symmetric relations $\id$, $\split$, $\cop$, and $\emptyset$,
which are defined as follows:

\begin{itemize}
\item For all $a, a' \in \AA$, $a \emptyset a'$ means that there is no
  relation between arcs $a$ and $a'$.
\item For all $a, a' \in \AA$, $a\ \id \ a'$ means that $a=a'$ (identity relation over $\AA)$.
\item For all $l, l', l_1, l_2 \in \LL$, $[l;l_1] \split [l';l_2]$ means that there is a split of emergy at node $l$ if $l=l'$ (node $l$ is called a split); else, it means that $l$ and $l'$ are emergy sources.
\item For all $l, l', l'' \in \LL$, $[l;l'] \cop [l;l'']$ means that there is a co-product at node $l$. Node $l$ is called a co-product.
\end{itemize}
For the example of Figure \ref{fig:example}, we have $[1;3] \split [2;10]$ because 1 and 2 are sources. Because 3, 5, 6, 7 and 10 are splits we have $[3;4] \split [3;5]$, $[6;8] \split [6;9]$, $[7;9] \split [7;10]$ and $[10;4] \split [10;11]$. Finally, we have $[4;6] \cop [4;7]$ since node 4 is a co-product.

The relations $\emptyset, \id, \bot$ and $\cop$ satisfy 7 axioms
\cite[Section 3.1]{kn:Lecorre2012b}. These axioms are mainly used to
prove that the path computation algorithm in \cite[Section 4]{kn:Lecorre2012b}
begins and terminates. In this paper, only the last axiom is of importance:

\begin{itemize}
\item[(H0)]By convention, each source of the emergy graph is connected to
  only one node of the emergy graph $G$.
\end{itemize}

We denote
$$
\FF \pardef  \LL \times \LL = \{[l_1 ; l_2] : l_1 \in \LL, l_2 \in \LL \},
$$

and we base the modelling of the emergy circulation within the graph $G$
on the idempotent semiring (which is a formal language):

$$
F\pardef( \FF^*_{\zero},\cup,\bullet,\zero,\un),
$$
where
\begin{enumerate}
\item $\FF^*_{\zero} \pardef \FF^* \cup \zero$.
\item $\FF^*$ is the set of all words of finite length constructed
  over the alphabet $\FF$.
\item $\zero$ is the empty set. 

\item $\bigcup$ is the union of two words, which can be identified with the union if a word $m$ is identified with the set $\{m\}$. It means that $\FF^*_{\zero}$ is identified with the set of all parts of $\FF^*_{\zero}$, which is denoted by $2^{\FF^*_{\zero}}$.
\item $\bullet$ is the concatenation of two words, which is defined as follows:
$$
\begin{tabular}{cccc}
$\bullet :$ & $\FF^*_{\zero} \times \FF^*_{\zero}$ & $\rightarrow$ & $\FF^*_{\zero}$\\
& $(m,m')$ & $\mapsto$ & $m \bullet m'$
\end{tabular}
$$
The word $m\bullet m’$ is the new word obtained by joining the letters
of $m$ and the letters of $m'$ end-to-end. When there is no ambiguity,
the concatenated word $m \bullet m’$ will be denoted by $mm’$.

\item $\un$ is the empty word.



\end{enumerate}

In \cite{kn:Lecorre2012b} formal language theory was used to compute 
relevant words by rewriting systems which are easy to program. 

But, from now till the end of the paper the graph theory and its 
vocabulary is applied.

\subsection{The emergy path}
\label{subEmpath}

Let us consider an emergy graph $G$, where $G=(\LL, \SS, \II, \OO, F, \AA, \id, 
\split, \cop, \emptyset)$. The definition of emergy path is based on 
the formal language $F$. The reader must be aware of the following:

\begin{rem}
\
\begin{itemize}
\item An arc $[l_1;l_2]$ is a letter of the language $F$.

\item A path in the graph $G$ is a sequence of consecutive arcs in $G$. It is 
thus a particular word of the language $F$. 
\end{itemize}
\end{rem}

Because of assumption (A2) the emergy evaluation is based on
considering particular paths in the emergy graph called 
emergy paths (see Definition~\ref{defPath}).

In the context of emergy we have:

\begin{itemize}
  \item the union of two paths $\pi$ and $\pi'$,
$\pi \cup \pi'$, models the fact that emergy can flow through
$\pi$ or $\pi'$ or both.

  \item the concatenation of two paths $\pi$ and $\pi'$,
$\pi \bullet \pi'$, models the fact that emergy can flow through
    $\pi$ and then can flow through $\pi'$. It is understood that
    $\pi \pi'$ is again a path.

  \item The path $\zero$ models the fact that emergy cannot
    circulate. From this modelling we deduce that:

    \begin{itemize}
    \item $\zero$ is absorbing for $\bullet$, i.e. for all path $\pi$:
      \[
\zero \pi = \pi \zero= \zero.
\]
Which means that if the emergy cannot circulate from upstream or downstream
a path $\pi$ then it cannot circulate on the whole path $\zero \pi$ or
$\pi \zero$.
\item $\zero$ is the neutral element for $\cup$, i.e. for all path $\pi$:
  \[
\pi \cup \zero = \zero \cup \pi = \pi.
\]
Which may be interpreted as follows. If the emergy has the choice (modelled by
operator $\cup$) between: impossibility to circulate (modelled by $\zero$)
or possibility to circulate through the path $\pi$ then emergy circulates
through $\pi$. If $\pi=\zero$ then emergy cannot circulate and we
have: $\zero \cup \zero = \zero$.
      \end{itemize}

\item The path $\un$ is the empty path. It has no physical meaning. 
It satisfies for all path $\pi$: $\un \pi = \pi \un =\pi$. And 
$\un \; \un = \un$.
    
    \end{itemize}

\begin{defi}
\label{defPath}
\ 
\begin{description}
\item[\textit{Path:}] it is a sequence of consecutive arcs.
Formally, it is an element $\pi$ of set  $\FF^*_{\zero}$ which has the form $\pi=\zero$ or $\pi=\un$ or $\pi=[l_1;l_2][l_2;l_3]\cdots[l_{k-2};l_{k-1}]$ $[l_{k-1};l_k]$, with $l_j \in \LL$, for $1 \leq j \leq k$ and $k \ge 2$. 

\item[\textit{Emergy path:}] it is a path which starts from a source and has no repeated node, except the last one which can be repeated once, ie. a path $\pi=[l_1;l_2][l_2;l_3]\cdots[l_{k-2};l_{k-1}] [l_{k-1};l_k]$ such that $l_1\in \SS$, $l_j \in \II \cup \OO$ for $2 \leq j \leq k$, and $l_i \neq l_j$ for $1\leq i < j \leq k-1$. 

\item[\textit{Path length:}] the length $lg(\pi)$ of a path $\pi$ is equal to $-\infty$ if $\pi=\zero$, and is equal to $0$ if $\pi=\un$; otherwise the length of $\pi$ is equal to the number of arcs $[l_j;l_{j+1}]$ which compose the path. 

\end{description}

\end{defi}

\begin{ex*}[continued]
Path $[1;3][3;5][5;7][7;10][10;4][4;6][6;8]$ is an emergy path since it starts form a source and has no repeated node. Path $[1;3][3;4][4;7]$ $[7;10][10;4]$ is also an emergy path since the only repeated node is the last one.
On the contrary, path $[1;3][3;4][4;7][7;10][10;4][4;6]$ is not an emergy path since node 4 is repeated but is not the last node. 
\end{ex*}

\subsection{Emergy as a max-plus linear function: the axiomatic basis}
\label{subEmFct}

When flows of emergy derive from different sources the emergies of the flows must be added \cite[p. 92]{kn:Odum96}: in the following, it is modelled by Axiom $(\varphi.4.1)$ (see Definition \ref{def:axioms}).

When flows of emergy derive from a same source and are joined at a node there are two cases to compute the emergy downstream this node: if the flows are of the same kind the emergies of the flows must be added \cite[p. 92]{kn:Odum96}, else the maximum of emergies of the flows must be taken \cite[Fig. 3.7 p. 51, p. 92]{kn:Odum96}. This is modelled by Axioms $(\varphi.4.2)$ and $(\varphi.4.3)$, respectively (see Definition \ref{def:axioms})

Recalling that a split node divides into emergies of the same kind and a co-product node divides into emergies of different kinds, it is noticed in \cite{kn:Lecorre2012b} that the emergy calculus is based on two main operators:

\begin{itemize}

\item The addition associated with split node

\item The maximum associated with co-product node

\end{itemize}
and the emergy flowing between two nodes of the emergy graph is
defined as a nonnegative-valued $(\max,+)$-linear function of emergy
paths and emergy sources. This kind of function is well-known in the
context of $(\max,+)$ algebra. The $(\max,+)$ algebra or tropical
algebra denotes the set of real numbers equipped with the $\max$
operator (which plays a role similar to the usual addition) and the
$+$ operator (which plays a role similar to the usual multiplication).
The interested reader by the vast litterature on this domain is
referred to e.g. \cite{kn:Bac-cooq}, \cite{kn:Golan99},
\cite{kn:Glazek02}, \cite{kn:KoloMaslov97}, \cite{kn:RGetal05}.

Let $G =(\LL, \SS, \II, \OO, F, \AA, \id, \split, \cop, \emptyset)$ be
an emergy graph (see Section \ref{subsec:emgraph}). We shall use the following notations:
\begin{nota}
\label{nota:nota}
\
\begin{description}
\item[$\pmb{Succ(i)}$:] the set of immediate successors of node $i$, i.e. $Succ(i) \pardef \{j : [i;j] \in \AA \}$.

\item[$\pmb{succ(i)}$:] the unique immediate successor (see Axiom (H0))
 of the emergy source $i \in \SS$. 

\item[$\pmb{A([l;l'])}$:] for a set of paths $A$, it is the set of elements of $A$ which end by arc $[l;l']$. 


\item For a path $\pi$ and a node $i$ we define $\pi_i$ as follows: $\pi_i=\zero$ if $\pi$ does not contain node $i$; else $\pi$ is the sequence of arcs starting by the first occurence of node~$i$.

\item[$\pmb{A_i}$:] for a set of paths $A$, we have

$$
A_i \pardef \left\{
\begin{tabular}{ll}
$\emptyset$ & if no path of $A$ contains node $i$,\\
$\{\pi_i: \pi \in A \}$ & else.\\
\end{tabular}
\right.
$$

\item[$\pmb{A_i([l;l'])}$:] for a set of paths $A$, it is the set $\{\pi_i: \pi \in A([l;l'])\}$.

\item[$\pmb{\empaths}$:] it is the set of all emergy paths.

\item[$\pmb{\cal P}$:] it is the set of paths obtained from emergy paths by removing first arcs until any node $i$, with $i \in \SS \cup \II$, i.e. 
$$
\pmb{\cal P} \pardef \bigcup_{i \in \SS \cup \II} \empaths_i.
$$

\item[$\pmb{\pi P}$:] for a path $\pi$ which ends by $i$ and a set of paths $P$ which start by $i$, it is  the set of paths obtained by the concatenation of $\pi$ and paths of $P$, i.e 
\[
\pi P \pardef \{\pi p : p \in P\}. 
\]
Note that if $\pi=\zero$ or $P=\emptyset$ we have $\pi P = \emptyset$, and if $\pi=\un$ we have $\pi P=P$.

\end{description}
\end{nota}

\begin{ex*}[continued]
Paths of $\empaths([9;13])$, i.e. emergy paths ending by $[9;13]$, are enumerated in Table 1. There are 6 emergy paths in $\empaths([9;13])$. 
Let $A=\{ \pi_1, \pi_2, \pi_4, \pi_5 \}$. We have $A_4 = \{ [4;6] [6;9] [9;13], [4;7] [7;9] [9;13] \}$ and $A_3([6;9]) = \{ [3;4] [4;6] [6;9], [3;5] [5;7] [7;10] [10;4][4;6][6;9]\}$.

Let $\pi=[1;3][3;5][5;7]$ and $P= \{[7;9] [9;13],  [7;10] [10;4][4;6][6;9][9;13] \}$. We have $\pi P = \{\pi_3, \pi_4 \}$.

\end{ex*}

\begin{table}
\centering
\begin{tabular}{c|l}
Path	& Description \\
\hline
$\pi_1$ &	$[1;3] [3;4] [4;6] [6;9] [9;13]$ \\
$\pi_2$ &	$[1;3] [3;4] [4;7] [7;9] [9;13]$ \\
$\pi_3$ &	$[1;3] [3;5] [5;7] [7;9] [9;13]$ \\
$\pi_4$ &	$[1;3] [3;5] [5;7] [7;10] [10;4][4;6][6;9][9;13]$ \\
$\pi_5$ &	$[2;10] [10;4] [4;6] [6;9] [9;13]$ \\
$\pi_6$ &	$[2;10] [10;4] [4;7] [7;9] [9;13]$
\end{tabular}
\caption{Paths of $\empaths([9;13])$, i.e. emergy paths ending by arc $[9;13]$.}
\label{tab:paths}
\end{table}

From now on, it is assumed that the set of all emergy paths $\empaths$
is given.  There exist several algorithms for computing this set: see
for example \cite{kn:Tennenbaum88} (track summing method),
\cite{kn:Marvugliaetal2011,kn:Scale2013} (graph search), or
\cite{kn:Lecorre2012b} (rewriting system theory).

Let us introduce the following functions. Let $\RR^+$ be the
set of nonnegative reals.

\begin{itemize}
\item The emergy function $\theta : \LL \rightarrow \RR^+$ such 
that $\theta(l_1)$ is the emergy of the source $l_1$, if $l_1 \in \SS$ 
and $0$ otherwise.

\item The weight function $\omega : \LL \times \LL \rightarrow [0,1]$ 
such that $\omega([l_1;l_2])$ corresponds to the pourcentage of emergy 
circulating on $[l_1;l_2]$ if $[l_1;l_2] \in \AA$ and $0$ otherwise.
\end{itemize}

Finally, let us define the max-plus linear function $\varphi$ which
allows us to compute the emergy flowing on every arc of the emergy
graph $G$.

Note that the definition of $\varphi$ is borrowed from
\cite[subsection 3.3]{kn:Lecorre2012b} and restricted to $2^\Setpaths$ which
induces to separate the cases of sources and split nodes. Thus,
formulation of \cite[axiom $(\varphi.4.1)$]{kn:Lecorre2012b} is
replaced with two axioms, so the axiomatic basis remains unchanged.

\begin{figure}
\centering
\includegraphics[scale=0.5]{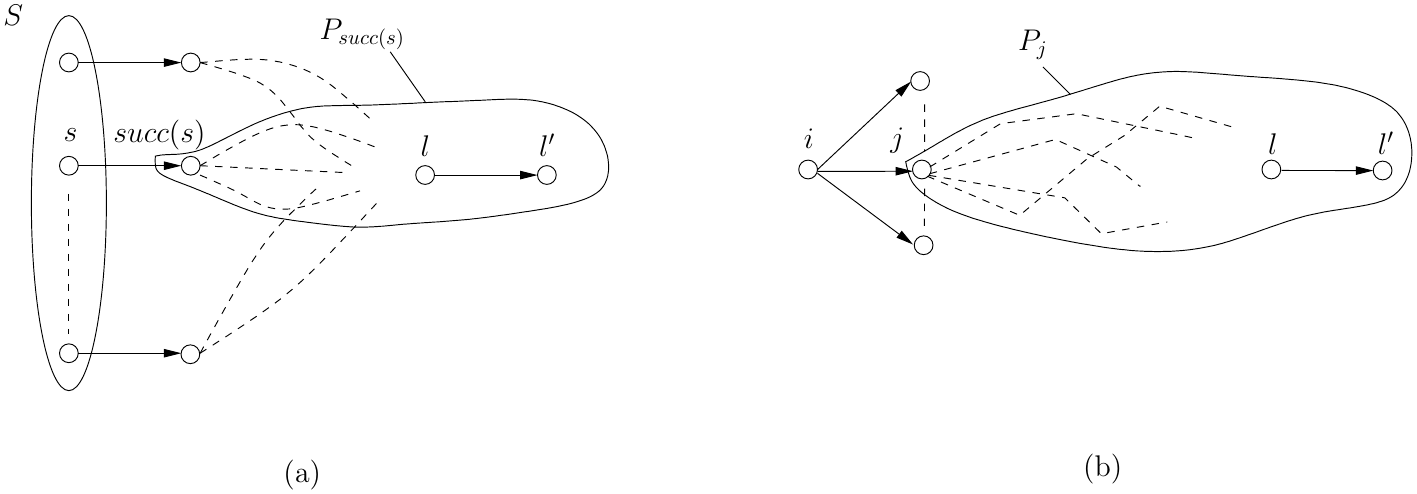}
\caption{Set of paths considered in Axioms ($\varphi.4.1$)-($\varphi.4.3$).}
\label{fig:axioms}
\end{figure}

\begin{defi}[Auxiliary function $\varphi$, \cite{kn:Lecorre2012b}]
\label{def:axioms}
The set function $\varphi: 2^\Setpaths \rightarrow \RR^+$ satisfies the following axioms:

\begin{description}
\item[$(\varphi.0)$:] $\varphi(\un) =1$, $\varphi(\zero)=0$ and $\varphi(\emptyset)=0$.
\item[$(\varphi.1)$:] $\forall \pi \in \Setpaths, \varphi(\pi) = \varphi(\{\pi\})$.
\item[$(\varphi.2)$:] $\varphi([l;l']) = \left\{
\begin{tabular}{ll}
$\omega([l;l'])$ & if $l,l' \notin \SS$,\\
$\theta(l) \omega([l;l'])$ & if $l \in \SS$ and $l' \notin \SS$,\\
\end{tabular}
\right.$
\item[$(\varphi.3)$:] $\forall \pi \in \Setpaths$, $\forall P \subseteq \Setpaths$, $\varphi(\pi P)=\varphi(\pi)\varphi(P)$.

\item[$(\varphi.4)$:] Let $P \subseteq \Setpaths$:

\begin{description}
\item[$(\varphi.4.1)$:] 
  If paths of $P$ start from a set of sources $S$, i.e. $S \subseteq \SS$ (see (a) of Figure \ref{fig:axioms})

$$
\varphi(P) = \sum_{s \in S} \varphi([s;succ(s)] P_{succ(s)}).
$$

\item[$(\varphi.4.2)$:] 
  If paths of $P$ start from a split $i$ (see (b) of Figure \ref{fig:axioms})

$$
\varphi(P) = \sum_{j \in Succ(i)} \varphi([i;j] P_j).
$$

\item[$(\varphi.4.3)$:] 
  If paths of $P$ start from a co-product $i$ (see (b) of Figure \ref{fig:axioms})

$$
\varphi(P) = \max_{j \in Succ(i)} \varphi ([i;j] P_j).
$$
\end{description}

\end{description}
\end{defi}

Now, we are able to recall the definition of the emergy measure flowing on an arc of an emergy graph.

\begin{defi}[Emergy evaluation, \cite{kn:Lecorre2012b}]
\label{def:emergy}
Let us consider the emergy graph $G$.
The emergy flowing on arc $[l;l']$, where $l,l' \in \LL$, is defined by
$$
\label{eq:emergy}
Em([l;l']) \pardef \varphi(\empaths([l;l'])).
$$
\end{defi}

\section{A mathematical formulation of Odum’s Maximum Empower Principle}
\label{secMEP}
Let us consider an emergy graph $G$, where $G=(\LL, \SS, \II, \OO, F, \AA, \id, 
\split, \cop, \emptyset)$  (see section \ref{subsec:emgraph}). We recall that the set of emergy paths $\empaths$ (see Definition \ref{defPath}) is assumed to be given. We mainly use Notations \ref{nota:nota}, so $\empaths([l;l'])$ denotes the set of all emergy paths ending by arc $[l;l']$.
The emergy function $\theta$ and the weight function $\omega$ of $G$ (see Section \ref{subEmFct}) are assumed to be known.
The emergy flowing on arc $[l;l']$ of $G$ is defined as $\varphi(\empaths([l;l'])$ (see Definition \ref{def:emergy}) where 
$\varphi$ is the auxiliary function of Definition \ref{def:axioms}.

We shall mainly use two kinds of path decomposition in our proofs:
\begin{rem}
\label{rem:sources}
Let $A \subseteq \empaths$.
\begin{enumerate}
\item Since every path of $A$ starts by a source $s$ of $\SS$ we have 
$$
A=\bigcup_{s \in \SS} A_s
$$
\item For $i \in \LL$ we have 
$$
A_i=\bigcup_{j \in Succ(i)} [i;j]A_j
$$
\end{enumerate}
\end{rem}


We shall also use the fact that emergy of a set of emergy paths can be decomposed source by source:
\begin{propo}[Decomposition principle]
\label{prop:decomposition}
If $A$ is a subset of $\empaths([l;l'])$ then 
$$
\varphi(A)=\sum_{s \in \SS} \varphi(A_s).
$$
\end{propo}
\proof
By Remark \ref{rem:sources} we have $A=\bigcup_{s \in \SS} A_s$.
By Axiom (H0) every source $s$ is connected to only one node, so we have  $A_s=[s;succ(s)]A_{succ(s)}$, hence $A=\bigcup_{s \in \SS} [s;succ(s)]A_{succ(s)}$.
Since paths of $A$ starts from sources and end by the same arc, Axiom ($\varphi.4.1$) applies to $\varphi(A)$, so $\varphi(A)= \sum_{s \in \SS} \varphi([s;succ(s)] A_{succ(s)})$, i.e. $\varphi(A)= \sum_{s \in \SS} \varphi(A_s)$.~\cqfd

We introduce the \textit{compatibility} binary relation on $\Setpaths$ (see Notation \ref{nota:nota}). If two paths are compatible, emergy flows on both paths. On the contrary, if they are not, emergy can flow only on one of the two paths (the one that maximizes emergy). This is due to the fact that at a co-product node only one path can be used by the emergy. 
It is defined as follows.

\begin{defi}[Compatible paths]
\label{def:compatible}
Paths $\pi$ and $\pi'$ of $\Setpaths$, where $\pi=[l_1;l_2][l_2;l_3] \cdots [l;l']$ and $\pi'=[l'_1;l'_2][l'_2;l'_3] \cdots [l;l']$, are compatible relatively to arc $[l;l']$ if one of the following cases occurs:
\begin{enumerate}
\item $\pi = \pi'$,
\item $l_1$ and $l'_1$ are different sources,
\item $l_k$ is a split, where $k \ge 1$ and $l_i = l'_i$, for $1 \leq i \leq k$.
\end{enumerate}
\end{defi}

\begin{rem}
\label{rem:coproduct}
It is important to notice that paths $\pi$ and $\pi'$ are not compatible if $l_k$ is a co-product, where $k \ge 1$ and $l_i = l'_i$, for $1 \leq i \leq k$.
\end{rem}

In our approach we consider an emergy graph as a dynamical system (in steady-state, recall Assumption (A0)). Hence, we use a vocabulary borrowed from dynamical systems theory and define the notion of \textit{emergy state} (see Definition \ref{def:emergyState}) and \textit{emergy attractor} (see Definition \ref{def:attractor}). As a consequence we have the correspondance between ecological theory and dynamic systems theory:

\begin{table}
\centering
\begin{tabular}{ccc}
Ecological theory & & Dynamic systems theory \\
\hline
network & $\Leftrightarrow$ & emergy graph \\
network design & $\Leftrightarrow$ & emergy state  \\ 
sustainable design &  $\Leftrightarrow$ & emergy attractor
\end{tabular}
\caption{The correspondance used for proving the MEP.}
\label{tab:correspondance}
\end{table}

\begin{defi}
\
\begin{description}
\label{def:emergyState}
\item[Emergy state:]
an emergy state, relatively to an arc $[l;l']$, is a set of pairwise compatible emergy paths which end by arc $[l;l']$. It is a subset of $\empaths([l;l'])$.

\item[Emergy state set:] 
the set of all emergy states, relatively to an arc $[l;l']$, is denoted by $\Setemstates([l;l'])$. Note that $\Setemstates([l;l'])$ is a part of $2^{\empaths([l;l'])}$. 
\end{description}
\end{defi}

\begin{ex*}[continued]
Set $\{\pi_1, \pi_5, \pi_6\}$ is an emergy state relatively to $[9;13]$ since the three paths are pairwise compatible (see Table \ref{tab:compatible}). On the contrary, the set $\{\pi_1, \pi_2, \pi_6\}$ is not an emergy state relatively to $[9;13]$ because paths $\pi_1$ and $\pi_2$ divide at node 4 which is a co-product node. 
Note that there are at most $2^6$ elements in $\Setemstates([9;13])$ since $\empaths([9;13])$ contains 6 emergy paths.
\end{ex*}

We introduce function $\fhat$ which is a filter that removes incompatible paths from a set of paths: given a set of paths $A$, that end by the same arc and originate from the same node, it returns a set of compatible paths $\fhat(A)$ such that $\fhat(A) \subseteq A$ and $\varphi(\fhat(A)) = \varphi(A)$.

\begin{defi}[Filter function]
Function $\fhat$, where $\fhat: 2^\Setpaths \rightarrow 2^\Setpaths$, is such that, for $A \subseteq {\cal P}_i([l;l'])$ and $i \in \SS \cup \II$,

$$
\fhat(A) = \left\{
\begin{tabular}{ll}
$\{[l;l']\}$ & if $i=l$,\\
$\bigcup_{j \in Succ(i)} [i;j] \fhat(A_j)$ & if $i$ is a split or a source,\\
$[i;j^*] \fhat(A_{j^*})$ & if  $i$ is a co-product. \\
where $j^* \in \argmax_{j \in Succ(i)} \varphi([i;j] \fhat(A_j))$ & \\
\end{tabular}
\right.
$$

\end{defi}

\begin{propo}
If $A$ is a subset of ${\cal P}_i([l;l'])$, where $i \in \SS \cup \II$, then 
\begin{subequations}

\begin{equation}
\label{subeq:fhat1}
\mbox{paths of } \fhat(A) \mbox{ are pairwise compatible},
\end{equation}
\begin{equation}
\label{subeq:fhat2}
\varphi(\fhat(A))=\sum_{\pi \in \fhat(A)} \varphi(\pi),
\end{equation}
\begin{equation}
\label{subeq:fhat3}
\varphi(\fhat(A)) = \varphi(A).
\end{equation}
\end{subequations}
\end{propo}
\proof
We give a proof by induction on the maximum length of a path of $A$. If $\max_{\pi \in A} lg(\pi)=1$ then $i=l$ and $\fhat(A)$ returns $\{[l;l']\}$: since $A=\{[l;l']\}$, \eqref{subeq:fhat1}-\eqref{subeq:fhat3} are true. Else, assume that $\fhat(A)$ returns a set that verifies \eqref{subeq:fhat1}-\eqref{subeq:fhat3} when $1\leq \max_{\pi \in A} lg(\pi) \leq n$, and consider a set $A$ such that $\max_{\pi \in A} lg(\pi)=n+1$. By Remark \ref{rem:sources} we have $A = \bigcup_{j \in Succ(i)} [i;j]A_j$, so $\max_{\pi \in A_j} lg(\pi)=n$ for $j \in Succ(i)$.

\begin{enumerate}
\item Property \eqref{subeq:fhat1}: let $j \in  Succ(i)$, and let $\pi_1$ and $\pi_2$ be two distinct paths of $\fhat(A_j)$.
Since paths of $\fhat(A_j)$ are compatible (by induction hypothesis) either $j$ is a split or $\pi_1$ and $\pi_2$ have a common path which ends by a split (case 3 of Definition \ref{def:compatible}). Hence,  $[i;j]\pi_1$ and $[i;j]\pi_2$ are compatible, so paths of $[i;j]\fhat(A_j)$ are pairwise compatible.
If there $i$ is a source or a co-product, there is only one successor $j$ of $i$ such that $\fhat(A_j) \neq \emptyset$ and the property is true. If $i$ is a split, consider two successors $j_1$ and $j_2$ of $i$, and let $\pi_1 \in \fhat(A_{j_1})$ and $\pi_2 \in \fhat(A_{j_2})$. Since case 3 of Definition \ref{def:compatible} applies, paths $[i;j_1]\pi_1$ and $[i;j_2]\pi_2$ are compatible, so paths of $\bigcup_{i \in Succ(i)} [i;j]\fhat(A_j)$, i.e. of $\fhat(A)$, are pairwise compatible. Thus, Property \eqref{subeq:fhat1} is true.

\item Properties \eqref{subeq:fhat2} and \eqref{subeq:fhat3}: 
\begin{itemize}
\item If $i$ is a source or a co-product then let $j$ be the single successor of $i$. We have $A =[i;j] A_j$ and $\fhat(A) =[i;j]\fhat(A_j)$, so $\varphi(\fhat(A)) = \varphi([i;j]\fhat(A_j))$. We get, by Axiom ($\varphi.3$) with $\pi=[i;j]$ and $P=\fhat(A_j))$, 
$$
\varphi(\fhat(A)) = \varphi([i;j]) \varphi(\fhat(A_j)).
$$
By induction hypothesis, $\varphi(\fhat(A_j)) = \sum_{\pi \in \fhat(A_j)} \varphi(\pi)$ and $\varphi(\fhat(A_j)) = \varphi(A_j)$, which implies respectively 

\begin{center}
\begin{tabular}{llll}
$\varphi(\fhat(A))$ & =  & $\sum_{\pi \in \fhat(A_j)} \varphi([i;j]) \varphi(\pi)$ & \\
 & = & $\sum_{\pi \in \fhat(A_j)} \varphi([i;j] \pi)$  & (by Axiom ($\varphi.3$) )\\
& = & $\sum_{\pi \in \fhat(A)} \varphi(\pi)$, & 
\end{tabular}
\end{center}
and
\begin{center}
\begin{tabular}{llll}
$\varphi(\fhat(A))$ & =  & $\varphi([i;j]) \varphi(A_j)$ & \\
 & = & $\varphi([i;j] A_j)$ & (by Axiom ($\varphi.3$)) \\
& = & $\varphi(A),$ & 
\end{tabular}
\end{center}
Hence, Properties \eqref{subeq:fhat3} and \eqref{subeq:fhat2} are true when $i$ is a source or a co-product.

\item If $i$ is a split then $\fhat(A) = \bigcup_{j \in Succ(i)} [i;j]\fhat(A_j)$. By Axiom ($\varphi.4.2$) we get
$$
\varphi(\fhat(A)) = \sum_{j \in Succ(i)} \varphi([i;j]\fhat(A_j)).
$$
By Axiom ($\varphi.3$) with $\pi=[i;j]$ and $P=\fhat(A_j)$, we have
$$
\varphi(\fhat(A)) = \sum_{j \in Succ(i)} \varphi([i;j]) \varphi(\fhat(A_j)).
$$
By induction hypothesis, $\varphi(\fhat(A_j)) = \sum_{\pi \in \fhat(A_j)} \varphi(\pi)$ and $\varphi(\fhat(A_j)) = \varphi(A_j)$, which implies respectively 

\begin{center}
\begin{tabular}{llll}
$\varphi(\fhat(A))$ &  = & $\sum_{j \in Succ(i)} \varphi([i;j]) \sum_{\pi \in \fhat(A_j)} \varphi(\pi)$ & \\
 & = & $\sum_{j \in Succ(i)} \sum_{\pi \in \fhat(A_j)} \varphi([i;j]) \varphi(\pi)$ & (by distributivity) \\
 & = & $\sum_{j \in Succ(i)} \sum_{\pi \in \fhat(A_j)} \varphi([i;j] \pi)$ & (by Axiom ($\varphi.3$)) \\
 & = & $\sum_{\pi \in \bigcup_{j \in Succ(i)} [i;j]\fhat(A_j)} \varphi(\pi)$  & (by path decomposition) \\
 & = & $\sum_{\pi \in \fhat(A)} \varphi(\pi)$ &
\end{tabular}
\end{center}
and
\begin{center}
\begin{tabular}{llll}
$\varphi(\fhat(A))$ & = & $\sum_{j \in Succ(i)} \varphi([i;j]) \varphi(A_j)$ & \\
 & = & $\sum_{j \in Succ(i)} \varphi([i;j]A_j))$ & (by Axiom ($\varphi.3$)) \\
 & = & $\varphi(\bigcup_{j \in Succ(i)} [i;j] A_j)$ & (by Axiom ($\varphi.4.2$)) \\
 & = & $\varphi(A)$ & 
\end{tabular}
\end{center}

which proves Properties \eqref{subeq:fhat2} and \eqref{subeq:fhat3} when $i$ is a split.
\end{itemize}
\end{enumerate}
Thus, the proof by induction is completed.
\cqfd

\begin{propo}
\label{prop:monotone}
Let $A \subseteq {\cal P}_i([l;l'])$, where $i \in \SS \cup \II$. Assuming that $A$ contains at least two paths, let $\pi$ be a path of $A$. We have
\begin{equation}
\label{eq:monotone}
\varphi(\fhat(A\backslash \{\pi\})) \leq \varphi(\fhat(A))
\end{equation}
\end{propo}

\proof
Since $A$ has at least two elements we cannot have $i = l$ (otherwise $A = \{[l;l']\}$).
Let $j'$ be the successor of $i$ in path $\pi$, i.e. $\pi=[i;j']\pi'$, where $\pi'$ is a path starting by node $j'$ and end by $[l;l']$. 
\begin{itemize}

\item If $i$ is a split or a source we have, by definition of $\fhat$, 
$$
\fhat(A) = \bigcup_{j \in Succ(i)} [i;j] \fhat(A_j)
$$
and
$$
\fhat(A\backslash \{\pi\}) = \bigcup_{j \in Succ(i), j \neq j'} [i;j] \fhat(A_j).
$$
By Axiom $(\varphi.4.2)$ we get
$$
\varphi(\fhat(A)) = \sum_{j \in Succ(i)} \varphi([i;j] \fhat(A_j))
$$
and
$$
\varphi(\fhat(A\backslash \{\pi\})) = \sum_{j \in Succ(i), j \neq j'} \varphi([i;j] \fhat(A_j)).
$$
Since $\varphi$ is nonnegative, inequality \eqref{eq:monotone} holds.
\item If $i$ is a co-product there are two cases:
\begin{enumerate}
\item If $j' \in \argmax_{j \in Succ(i)} \varphi([i;j] \fhat(A_j))$ then 
$$\varphi(\fhat(A)) = \varphi([i;j'] \fhat(A_{j'}))
$$
and
$$
\varphi(\fhat(A\backslash \{\pi\})))  = \varphi([i;j'']\fhat(A_{j''})), \mbox{ where } j'' \in Succ(i), j'' \neq j'.
$$
Since $\varphi([i;j'']\fhat(A_{j''})) \leq \varphi([i;j'] \fhat(A_{j'}))$, inequality \eqref{eq:monotone} holds.

\item $If j' \notin \argmax_{j \in Succ(i)} \varphi([i;j] \fhat(A_j))$ then there exists $j'' \in Succ(i)$ such that $j'' \neq j'$ and $j'' \in \argmax_{j \in Succ(i)} \varphi([i;j] \fhat(A_j))$.
Hence, we have
$$\varphi(\fhat(A)) = \varphi([i;j''] \fhat(A_{j''}))$$
and
$$
\varphi(\fhat(A\backslash \{\pi\})))  = \varphi([i;j'']\fhat(A_{j''})),
$$
so inequality \eqref{eq:monotone} holds.
\end{enumerate}

\end{itemize}
\cqfd

\begin{defi}[Emergy attractor]
\label{def:attractor}
An emergy attractor $\attractor$, for an arc $[l;l']$ of $\AA$, is an emergy state of $\Setemstates([l;l'])$ such that $\varphi(\attractor) = Em([l;l'])$.
\end{defi}

\begin{theo}
\label{theo:attractor}
There exists an emergy attractor for every arc $[l;l']$ of $\AA$.
\end{theo}
\proof
By definition, we have $Em([l;l')]=\varphi(\empaths([l;l'])$. By Property \ref{subeq:fhat2}, we get $Em([l;l')]=\varphi(\fhat(\empaths([l;l']))$. Hence, $\attractor = \fhat(\empaths([l;l'])$.

\cqfd

We are now able to state the main result of the paper, which interpretation is given in the conclusion:

\begin{theo}[Maximum empower]
\label{theo:maxempower}
An attractor $\attractor$ for an arc $[l;l']$ of $\AA$ satisfies
$$
\varphi(\attractor) = \max_{\emstate \in \Setemstates([l;l'])} \varphi(\emstate).
$$

\end{theo}
\proof
Let us consider an emergy state $\emstate$ of $\Setemstates([l;l'])$. 
Since $\forall s \in \SS, \emstate_s \subseteq \empaths_s([l;l'])$ we get $\forall s \in \SS, \varphi(\fhat(\emstate_s)) \leq \varphi(\fhat(\empaths_s([l;l'])))$ by Proposition \ref{prop:monotone}. By Property \eqref{subeq:fhat3} we obtain $\forall s \in \SS, \varphi(\emstate_s) \leq \varphi(\empaths_s([l;l']))$.
By the decomposition principle (Proposition \ref{prop:decomposition}) we have 
$
\varphi(\empaths([l;l']))= \sum_{s \in \SS} \varphi(\empaths_s([l;l'])) \mbox{ and }
\varphi(\emstate) = \sum_{s \in \SS} \varphi(\emstate_s)
$, hence $\varphi(\emstate) \leq Em([l;l'])$.
By Theorem \ref{theo:attractor}, there exists an attractor $\attractor$ for arc $[l;l']$, so $\varphi(\attractor) = Em([l;l'])$ and the result follows.
\cqfd

The emergy of an emergy path $\pi$ is $\varphi(\pi)$. 
The emergy of an emergy state is the sum of the emergies of its paths:

\begin{theo}
\label{theo:emstatecalculus}
If $\emstate$ is an emergy state of $\Setemstates([l;l'])$, with $[l,l'] \in \AA$, then
$$
\varphi(\emstate) = \sum_{\pi \in \emstate} \varphi(\pi).
$$
\end{theo}
\proof
Since $\emstate$ is an emergy state, any pair of paths of $\emstate$ are compatible, i.e. no paths can divide at a co-product node (recall case 3 of Definition \ref{def:compatible}). Hence, $\fhat$ applied to $\emstate$ does not remove any path, so $\fhat(\emstate) = \emstate$. Since $\varphi(\fhat(\emstate)) = \sum_{\pi \in \fhat(\emstate)} \varphi(\pi)$, we get $\varphi(\emstate) = \sum_{\pi \in \emstate} \varphi(\pi)$.
\cqfd

The emergy of an emergy path is computed as follows:
\begin{propo}
\label{prop:empath}
Let $\pi$ be an emergy path, where $\pi=[s;l_1][l_1;l_2]$ $\cdots[l_k;l_{k+1}]$. We have
$$
\varphi(\pi) = \theta(s) \omega([s;l_1]) \prod_{1\leq i \leq k} \omega([l_i;l_{i+1}]).
$$
\end{propo}
\proof
By Axiom ($\varphi.1$) we have $\varphi(\pi)=\varphi(\{\pi\})$.
We can write $\{\pi\} = [s;l_1]\{[l_1;l_2] $ $\cdots [l_k;l_{k+1}]\}$ (see Notations \ref{nota:nota}) so that we get $\varphi(\{\pi\}) = \varphi([s;l_1])$ $ \varphi(\{[l_1;l_2]$ $\cdots[l_k;l_{k+1}]\})$ by Axiom ($\varphi.3$). Repeating this reasoning we obtain $\varphi(\{\pi\})=\varphi([s;l_1]) \varphi([l_1;l_2]) \cdots $ $\varphi([l_k;l_{k+1}])$.
By Axiom ($\varphi.2$) we have $\varphi([s;l_1]) = \theta(s) \omega([s;l_1])$ and $\varphi([l_i;l_{i+1}]) = \omega([l_i;l_{i+1}])$, for $1\leq i \leq k$. Hence, the result follows.
\cqfd

\section{Numerical example}
\label{secExample}
We consider the example taken from \cite{kn:Lietal2010} (whose emergy graph is given in Figure \ref{fig:example}), where the emergy of the sources are $\theta(1)=1000$ and $\theta(2)=500$.
Let us compute the emergy flowing on arc $[9;13]$, i.e. $Em([9;13])$. 

\begin{table}
\centering
\begin{tabular}{c|cccccc}
& $\pi_1$ & $\pi_2$ & $\pi_3$ & $\pi_4$ & $\pi_5$ & $\pi_6$ \\
\hline
$\pi_1$ & T & F & T & T & T & T \\
$\pi_2$ & F & T & T & T & T & T \\
$\pi_3$ & T & T & T & T & T & T \\
$\pi_4$ & T & T & T & T & T & T \\
$\pi_5$ & T & F & T & T & T & F \\
$\pi_6$ & T & F & T & T & F & T
\end{tabular}
\caption{Compatibility relation between paths of $\empaths([9;13))$ ('T' means True, 'F' means False).}
\label{tab:compatible}
\end{table}  

By Theorem \ref{theo:attractor}, there exists an attractor $\attractor$ such that $Em([9;13])=\varphi(\attractor)$ and, by Theorem \ref{theo:maxempower}, $\varphi(\attractor) = \max_{\emstate \in \Setemstates([9;13])} \varphi(\emstate)$.

Here, $\Setemstates([9;13])$ contains at most $2^6$ emergy states because there are 6 emergy paths that end by arc $[9;13]$ (see Table \ref{tab:paths}). However, it is possible to avoid enumeration of the 64 sets by noticing that paths $\pi_3$ and $\pi_4$ are compatible with all other paths (see Table \ref{tab:compatible}). 
Therefore, an attractor $\attractor$ is of the form $\attractor = \{\pi_3,\pi_4\} \cup \emstate$ with $\emstate \subseteq \{\pi_1,\pi_2,\pi_5,\pi_6\}$, i.e.
$$
\varphi(\attractor) = \varphi(\{\pi_3,\pi_4\} \cup \emstate).
$$
By Theorem \ref{theo:emstatecalculus} we have 
$$\varphi(\{\pi_3,\pi_4\} \cup \emstate) = \varphi(\pi_3) + \varphi(\pi_4) + \varphi(\emstate),$$
so we get
$$\varphi(\attractor) = \max_{\emstate' \subseteq \{\pi_1,\pi_2,\pi_5,\pi_6\}} (\varphi(\pi_3) + \varphi(\pi_4) + \varphi(\emstate')),$$
 i.e. $\varphi(\attractor) = \varphi(\pi_3) + \varphi(\pi_4) + \max_{\emstate' \subseteq \{\pi_1,\pi_2,\pi_5,\pi_6\}} \varphi(\emstate')$. Hence, finding $\attractor$ reduces to finding $\emstate$ with $\varphi(\emstate) =  \max_{\emstate' \subseteq \{\pi_1,\pi_2,\pi_5,\pi_6\}} \varphi(\emstate')$. 

\begin{figure}
\centering
\includegraphics[scale=0.5]{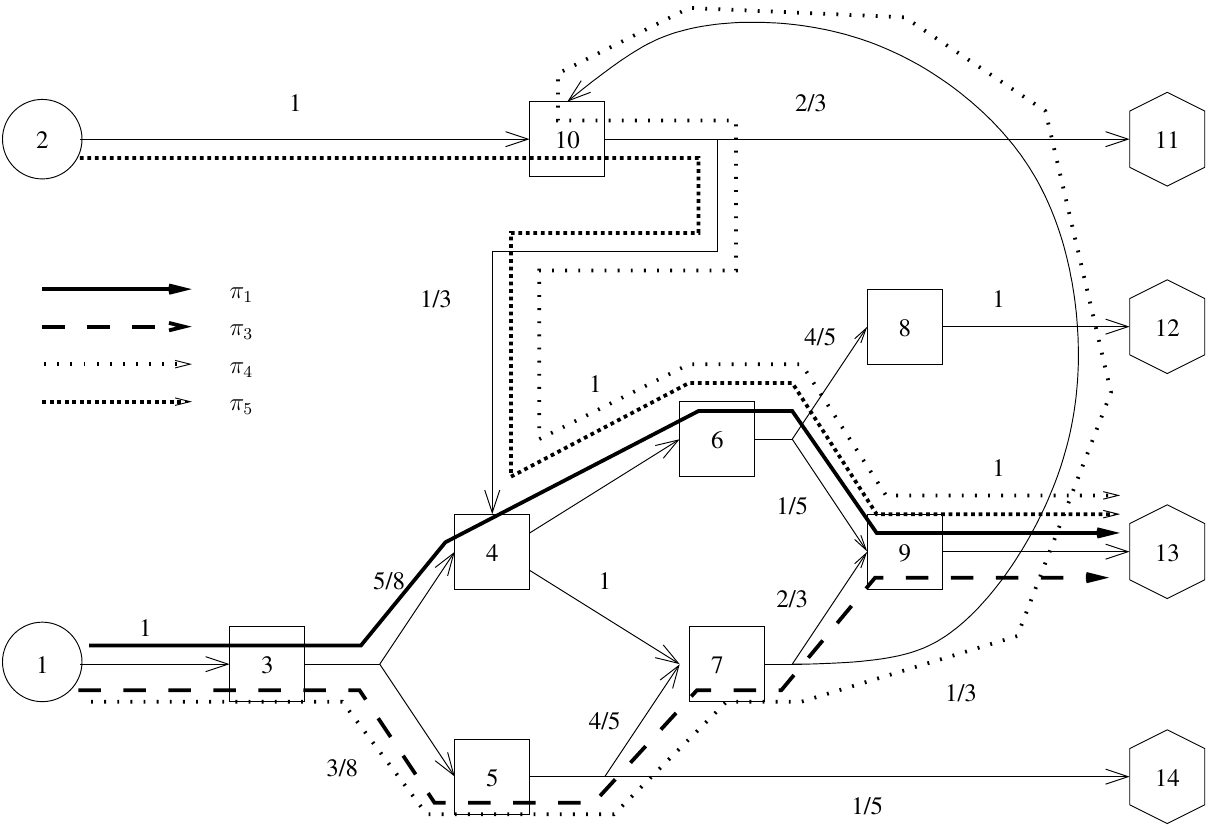}
\caption{The emergy paths of emergy state $\emstate_1$.}
\label{fig:emergyState1}
\end{figure}

\begin{figure}
\centering
\includegraphics[scale=0.5]{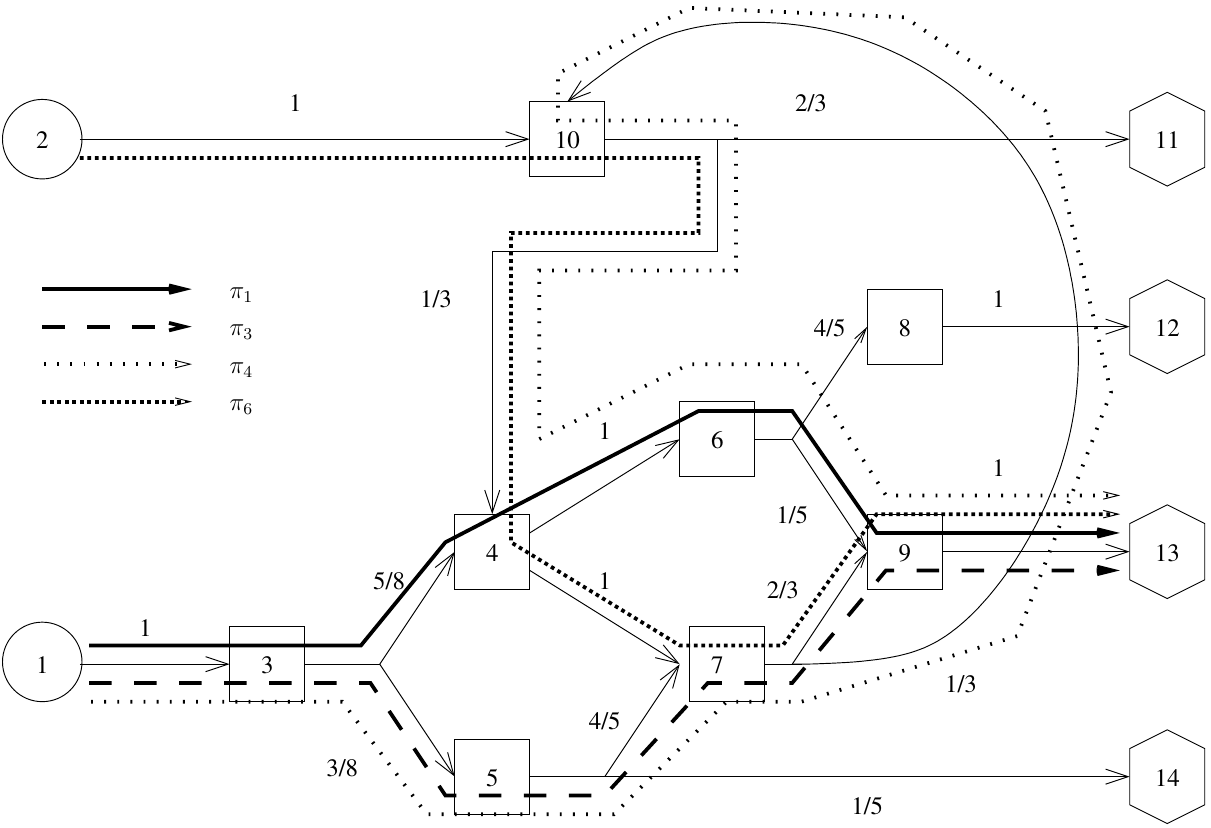}
\caption{The emergy paths of emergy state $\emstate_2$.}
\label{fig:emergyState2}
\end{figure}

\begin{figure}
\centering
\includegraphics[scale=0.5]{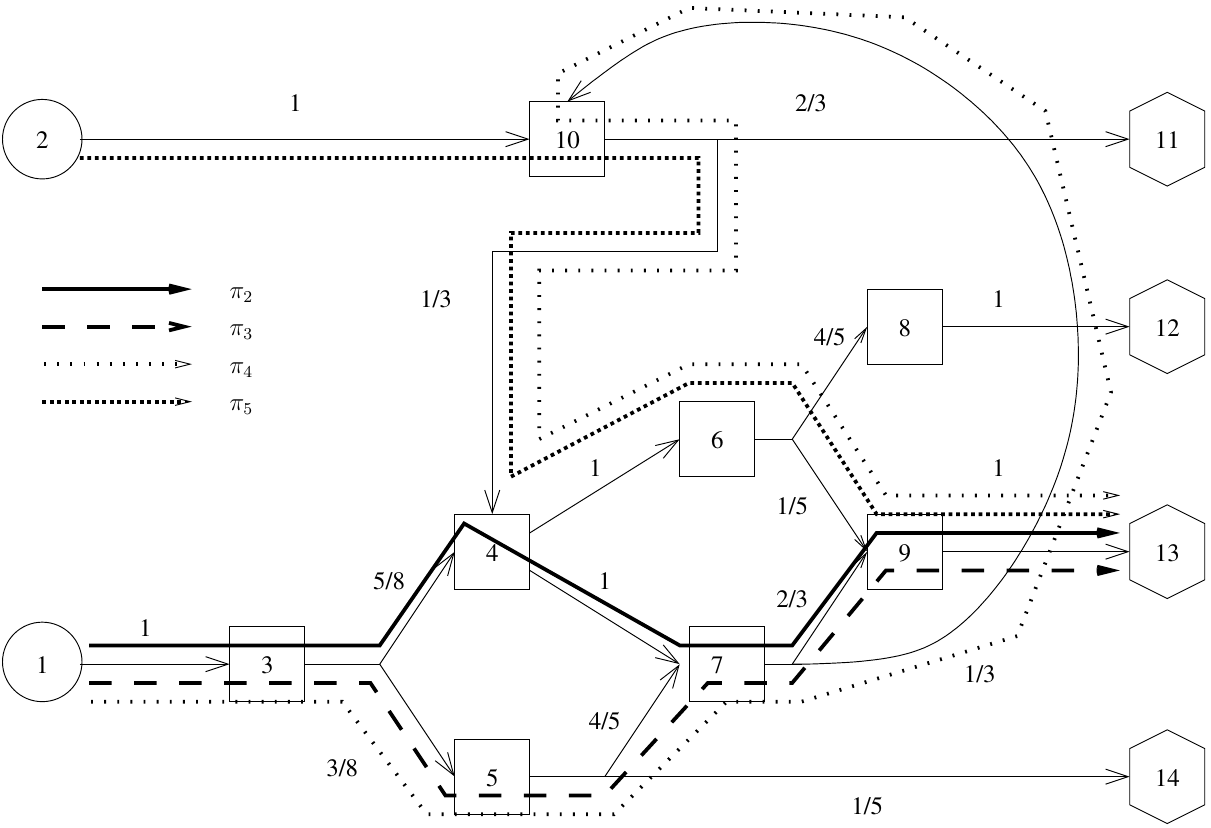}
\caption{The emergy paths of emergy state $\emstate_3$.}
\label{fig:emergyState3}
\end{figure}

\begin{figure}
\centering
\includegraphics[scale=0.5]{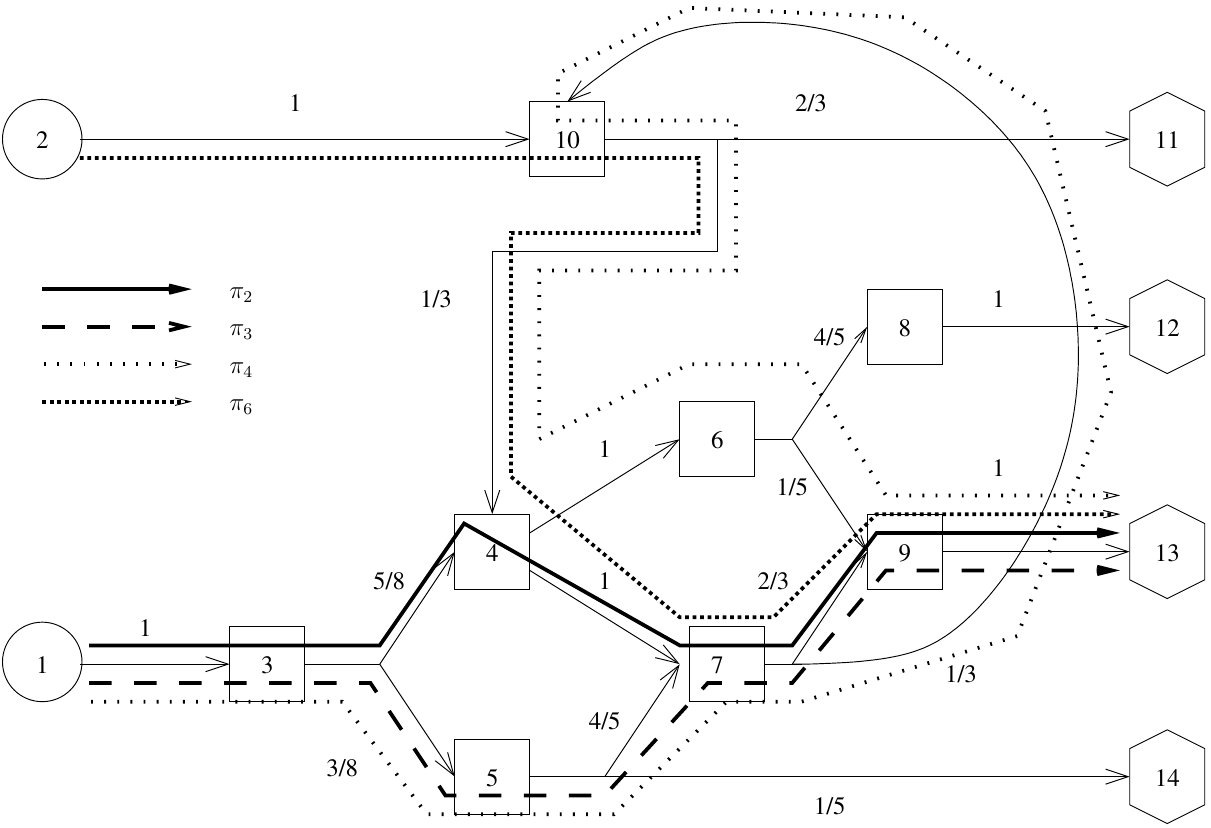}
\caption{The emergy paths of emergy state $\emstate_4$.}
\label{fig:emergyState4}
\end{figure}

Now, let us notice that $\pi_1$ and $\pi_2$ (resp. $\pi_5$ and $\pi_6$) are not compatible. Thus, we only have to consider 4 candidates for $\emstate'$ :  $\emstate'_1=\{\pi_1,\pi_5\}$,  $\emstate'_2=\{\pi_1,\pi_6\}$, $\emstate'_3=\{\pi_2,\pi_5\}$ and $\emstate'_4=\{\pi_2,\pi_6\}$.
As a consequence, the four candidates for $\attractor$ are: $\{\pi_1,\pi_3,\pi_4,\pi_5\}$, $\{\pi_1,\pi_3,\pi_4,\pi_6\}$, $\{\pi_2,\pi_3,\pi_4,\pi_5\}$ and $\{\pi_2,\pi_3,\pi_4,\pi_6\}$ (see Figures \ref{fig:emergyState1}, \ref{fig:emergyState2}, \ref{fig:emergyState3} and \ref{fig:emergyState4} respectively). Hence, 
\begin{equation}
\label{eq:em1}
Em([l;l']) =  \varphi(\pi_3) + \varphi(\pi_4) + \max \{ \varphi(\emstate'_1), \varphi(\emstate'_2), \varphi(\emstate'_3), \varphi(\emstate'_4) \}.
\end{equation}
By Proposition \ref{prop:empath} we have:
\begin{itemize}
\item $\varphi(\pi_1) = \theta(1) \omega([3;4]) \omega([4;6]) \omega([6;9]) \omega([9;13]) = 1000 \cdot \frac{5}{8} \cdot 1 \cdot \frac{1}{5} \cdot 1 = 15.625$.

\item $\varphi(\pi_2) = \theta(1) \omega([3;4]) \omega([4;7])  \omega([7;9])  \omega([9;13]) = 1000 \cdot \frac{5}{8} \cdot 1 \cdot \frac{2}{3} \cdot 1 = 416.667$.

\item $\varphi(\pi_3) = \theta(1)  \omega([3;5])  \omega([5;7])  \omega([7;9]) \omega([9;13]) = 1000 \cdot \frac{3}{8} \cdot \frac{4}{5} \cdot \frac{2}{3} \cdot 1 = 200$.

\item 
$\varphi(\pi_4) = \theta(1)  \omega([3;5])  \omega([5;7])  \omega([7;10])  \omega([10;4])  \omega([4;6])  \omega([6;9]) \omega([9;13])$

\hskip 0.95cm  $ = 1000 \cdot \frac{3}{8} \cdot \frac{4}{5} \cdot \frac{1}{3} \cdot \frac{1}{3} \cdot 1 \cdot \frac{1}{5} \cdot 1 = 6.667$.

\item $\varphi(\pi_5) = \theta(2)  \omega([10;4])  \omega([4;6])  \omega([6;9])  \omega([9;13]) = 500 \cdot \frac{1}{3} \cdot 1 \cdot \frac{1}{5} \cdot 1 = 33.333$.

\item $\varphi(\pi_6) = \theta(2)  \omega([10;4])  \omega([4;7])  \omega([7;9])  \omega([9;13]) = 500 \cdot \frac{1}{3} \cdot 1 \cdot \frac{2}{3} \cdot 1 = 111.111$.

\end{itemize}

By Theorem \ref{theo:emstatecalculus} we get:

$\varphi(\emstate'_1)=\varphi(\pi_1) + \varphi(\pi_5) = 15.625 + 33.333 = 48.958$,

$\varphi(\emstate'_2)=\varphi(\pi_1) + \varphi(\pi_6) = 15.625 + 111.111 = 126.736$,

$\varphi(\emstate'_3)=\varphi(\pi_2) + \varphi(\pi_5) = 416.667 + 33.333 = 450.000$,

$\varphi(\emstate'_4)=\varphi(\pi_2) + \varphi(\pi_6) = 416.667 + 111.111 = 527.778$.

Therefore, an attractor for arc $[9;13]$ is obtained by $\attractor = \{\pi_3,\pi_4\} \cup \emstate'_4$ and we get 
$$
Em([9;13]) = \varphi(\pi_3) + \varphi(\pi_4) + \varphi(\emstate'_4) = 200 + 6.667 + 527.778 = 734.445.
$$

It is interesting to recall the expression obtained by the algorithm proposed in \cite{kn:Lecorre2012b}:
$$Em([9;13]) = \varphi(\pi_3) + \varphi(\pi_4) + \max \{ \varphi(\pi_1), \varphi(\pi_2) \} +  \max \{ \varphi(\pi_5), \varphi(\pi_6) \}.$$
Noticing that it can be rewritten as
$$
\begin{array}{rcl}
Em([9;13])  & = & \varphi(\pi_3) + \varphi(\pi_4) \\
 & & + \max \{ \varphi(\pi_1) + \varphi(\pi_5), \varphi(\pi_1) + \varphi(\pi_6), \varphi(\pi_2) + \varphi(\pi_5), \varphi(\pi_2) + \varphi(\pi_6) \} \\
  & = & \varphi(\pi_3) + \varphi(\pi_4) + \max \{ \varphi(\emstate'_1), \varphi(\emstate'_2), \varphi(\emstate'_3), \varphi(\emstate'_4) \},
\end{array}
$$
we retrieve \eqref{eq:em1}.

\section{Conclusion}
\label{secConclusion}
Recall that the maximum empower principle (MEP) was expressed by Odum as maximization of ecological network designs:

\textit{''In the competition among self-organizing processes, network designs that maximize empower will prevail''} \cite[p. 16]{kn:Odum96}.

In this work, we have proposed and used a correspondance between ecological theory and dynamic systems theory (see Table \ref{tab:correspondance}), so that the MEP can be stated as in Theorem \ref{theo:maxempower}:

\textit{"In the competition among self-organizing processes, emergy states that maximize empower will prevail"}.

Moreover, a network design that maximises empower is called sustainable design by Odum \cite[p. 279]{kn:Odum96}, which corresponds to emergy attractor in our settings.

A consequence of Theorem \ref{theo:maxempower} is that empower computation is a new combinatorial optimization problem. This gives the opportunity to tackle empower computation by using techniques from the rich literature on combinatorial optimization.


\bibliographystyle{plain}

\bibliography{ref_lt}

\end{document}